\documentclass[%
reprint,amsmath,amssymb,
aps,
prl,
showkeys
]{revtex4-2}
\usepackage{xcolor}
\usepackage{graphicx}
\usepackage{dcolumn}
\usepackage{bm}
\usepackage{chngcntr}
\newcommand{\ik}[1]{{\color{black} #1}}
\begin{document}

	\title{Dual Role for Heterogeneity in Dynamic Fracture}
	
	\author{Itamar Kolvin}\email{ikolvin@gatech.edu}
	\affiliation{School of Physics, Georgia Institute of Technology, 837 State St NW, Atlanta, 30332, GA, USA}
	
	\author{Mokhtar Adda-Bedia}
	\affiliation{Laboratoire de Physique, CNRS, ENS de Lyon, Universit\'e de Lyon, F-69364 Lyon,  France}

	\date{\today}
	
	\begin{abstract}
		 We approach the problem of heterogeneous dynamic fracture by considering spatiotemporal perturbations to planar crack fronts. Front propagation is governed by local energy balance between the elastic energy per unit area available to fracture, $G$, and the dissipation \ik{in} creating new surfaces. $G$ is known analytically as a perturbation series in the crack front fluctuation. \ik{For dissipation} that monotonically increases with the crack speed, we derive an equation of motion for crack fronts that is second-order accurate. In the linear order, heterogeneity does not change the net speed of fracture. In the second order, nonlinear interactions of the front and the heterogeneous landscape populate an intermediate-scale fluctuation spectrum. We find that, when \ik{dissipation} weakly grows with velocity, nonlinearities globally amplify dissipation and reduce the crack speed. Strong velocity dependence, however, mitigates toughening effects and may facilitate fracture.
	\end{abstract}
	
	\maketitle

	Heterogeneous materials that locally vary in their mechanical properties are everywhere in geology~\cite{bedfordFaultRockHeterogeneity2022,rayEarthquakeNucleationFaults2017,taufiqurrahmanDynamicsInteractionsDelays2023,gounonRuptureNucleationPeriodically2022,faureExperimentalEvidenceSeismic2024,garagashFractureMechanicsRateandstate2021}, biology~\cite{barthelatFractureMechanicsBiological2023}, and engineering~\cite{aminiCentrifugationIndexMatching2021,shaikeeaToughnessMechanicalMetamaterials2022,guptaToughDuctileArchitected2024a}. The bulk properties of heterogeneous media, such as the Young modulus, are well-represented by macroscopic averaging over the material microstructure~\cite{hashinElasticModuliHeterogeneous1962}. In contrast, the consequences of heterogeneity for rapid cracks are only beginning to be ascertained. To determine how heterogeneity controls energy dissipation in dynamic fracture, the three-dimensional crack evolution must be resolved ~\cite{willisDynamicWeightFunctions1995,Ramanathan.97,movchanPerturbationsPlaneCracks1998,Morrissey.00,norrisMultiplescalesApproachCrackfront2007,wangHowHidden3D2022,weiComplexityCrackFront2024,riceThreedimensionalPerturbationSolution1994}. 
	
	Heterogeneity may produce toughening effects. Quasi-static cracks experience local and transient arrests at tough asperities which impede the overall progress of fracture~\cite{rouxEffectiveToughnessHeterogeneous2003,demeryMicrostructuralFeaturesEffective2014,vasoyaStudyTensileFailure2014,chopinDepinningDynamicsCrack2018,albertiniEffectiveToughnessHeterogeneous2021,lebihainSizeEffectsToughening2023,sannerWhySoftContacts2024,gaoFirstOrderPerturbationAnalysis1989,bouchaudScalingPropertiesCracks1997,ponsonDynamicForcingCrack2024}.
	In bulk fracture, the formation of complex out-of-plane surface structure may be triggered by  inhomogeneities~\cite{phamGrowthCracksMixedmode2016,steinhardtHowMaterialHeterogeneity2022,lebihainEffectiveToughnessPeriodic2020,lubomirskyQuenchedDisorderInstability2024}. The resultant growth in fracture surface area is associated with increased energy dissipation~\cite{sharonEnergyDissipationDynamic1996,Sagy.06,lubomirskyQuenchedDisorderInstability2024,weiComplexityCrackFront2024}. Still, the impact of heterogeneity on \textit{dynamic} cracks remains an open question. Existing computational methods, including the spectral boundary integral method~\cite{rochDynamicCrackFrontDeformations2023}, phase field simulations~\cite{ponsHelicalCrackfrontInstability2010,henryFractographicAspectsCrack2013,Chen.15,lubomirskyQuenchedDisorderInstability2024} and atomistic models~\cite{mollerInfluenceCrackFront2015,buehlerModelingAtomisticDynamic2022}, predict 3D crack motion by obtaining the full elastic fields. The need to resolve 3D or 2D singular dynamic fields limits the accessible system sizes. Alternatively, the fields near a running crack can be analytically determined through a perturbative approach \ik{which predicts, among else, crack front wave propagation}~\cite{willisDynamicWeightFunctions1995,Ramanathan.97,movchanPerturbationsPlaneCracks1998,Morrissey.00,norrisMultiplescalesApproachCrackfront2007}. \ik{In the presence of heterogeneity, the first order correction to the global dissipation is zero. Tractable expressions for higher order corrections are then needed to evaluate the contribution of heterogeneity to toughness.}
	
	To make progress, we analytically approximated the elastic fields near dynamic planar cracks at the second-order in the crack front fluctuation~\cite{kolvinComprehensiveStudyNonlinear2024a}. The local balance between the elastic energy flux and dissipation dictate\ik{s} a one-dimensional equation of motion for the crack front. Solutions for cracks traversing heterogeneous media showed that nonlinear interactions generate front fluctuations at intermediate length and time scales. Beyond mere averaging, the \ik{nonlinear couplings of front fluctuations to heterogeneity} renormalize the global dissipation and speed of fracture. We predict slower crack speeds and increased dissipation when the fracture energy is weakly velocity-dependent. Conversely, strongly velocity-dependent fracture energy may facilitate sufficiently rapid fracture. \ik{The effect of heterogeneity on the overall dissipation in dynamic fracture is thus nonlinear in the leading order.}
    
	\textit{Crack front equation of motion.} Dynamic cracks are governed by \ik{energy} balance $G=\Gamma$, between the elastic energy per unit area flowing into the crack tip, $G$, and the dissipation per unit area tied to the creation of new surfaces, $\Gamma$~\cite{Ramanathan.97,wangHowHidden3D2022}. To investigate how dynamic cracks interact with inhomogeneities, \ik{we obtained} $G$ locally as a function of the crack front geometry, velocity, and history~\cite{kolvinComprehensiveStudyNonlinear2024a}. \ik{We then complement energy balance by modeling $\Gamma$ as a velocity-dependent and space-varying quantity.}
	
	\ik{We} consider a semi-infinite crack \ik{propagating at velocity $V$} in the $y=0$ plane of a linearly elastic solid subject to remote time-independent tensile stresses~\cite{Freund.90}. The \ik{$x$ position of the unperturbed} crack front is \ik{$h(z,t) = Vt$}. Ahead of the crack, the tensile stress in the fracture plane is singular $\sigma_{yy}\sim (x-h)^{-1/2}$. The energy released by the crack per unit area is $G=G_r g(V)$, where the equilibrium energy release rate, $G_r$, is determined by the remote stresses and $g(V)$ is a universal function that regulates the flow of energy into the crack tip \ik{and limits the crack velocity at the Rayleigh wave speed $c_R$}~\cite{Freund.90}. For \ik{perturbed} crack fronts whose $x$ position is $h(z,t) = Vt + f(z,t)$, energy balance becomes
	
	\begin{equation}\label{3d_energy_balance}
		G_r g(V_\perp)\left(1+H\left[\{f(z,t'); t'\leq t\}\right]\right) = \Gamma(x,z; V_\perp)\,.
	\end{equation}
	
	\ik{The dissipation $\Gamma$ depends on local material properties and on the normal velocity $V_\perp(z,t) = (V+\partial_t f)/\sqrt{1+(\partial_z f)^2}$}. \ik{The nonlinear functional $H$ depends on prior front configurations and} may be approximated by a perturbation expansion~\cite{willisDynamicWeightFunctions1995,Ramanathan.97,movchanPerturbationsPlaneCracks1998,norrisMultiplescalesApproachCrackfront2007,kolvinComprehensiveStudyNonlinear2024a}. Eq.~(\ref{3d_energy_balance}) implicitly determines the instantaneous normal velocity $V_\perp$ as a function of the front history and toughness heterogeneity. \ik{We maintain an analytical approach that keeps the dynamics predicted by Eq.~(\ref{3d_energy_balance}) reversible. However, to approximate irreversible crack propagation, we focus on instances in which local negative velocities are few and negligible.}
	
	The response of dynamic crack fronts to small perturbations is known~\cite{Ramanathan.97,Morrissey.00,norrisMultiplescalesApproachCrackfront2007}. Let us decompose the front fluctuation into Fourier components $\hat{f}(k,\omega) = \int\! \mathrm{d}z\, \mathrm{d}t\, e^{-ikz-i\omega t} f(z,t)$, where $k$ is the wavenumber and $\omega$ is the frequency of a component. For small $f$, Ramanathan and Fisher~\cite{Ramanathan.97} obtained $\widehat{G} = G_r g(V)(1 +\widehat{\delta G})$ where 
	\begin{equation}\label{linear_dG}
		\widehat{\delta G} = -2 |k|P_1(\omega/|k|;V,\nu) \hat{f} + \mathcal{O}(f^2)\,,
	\end{equation}
	where $\nu$ is the Poisson ratio and $P_1$ is an explicit function of its arguments~(Appendix A, Eq.~(\ref{P1}) ,~\cite{kolvinComprehensiveStudyNonlinear2024a}). Through Eq.~(\ref{3d_energy_balance}), the kernel $-2 |k| P_1$ defines the linear response of the crack front to spatial variations in $\Gamma$. Particularly, the zero mode $P_1( \omega/|k|=c_f)=0$  at $c_f  \equiv \xi(V/c_R,\nu)\sqrt{c_R^2-V^2}$, where $\xi\simeq 1$~\cite{Ramanathan.97,Morrissey.00}, \ik{predicts}  one-dimensional dispersion-less waves of speed $c_f$ that propagate along the crack front. The \ik{existence} of front waves was corroborated numerically and experimentally~\cite{sharonPropagatingSolitaryWaves2001,fekakCrackFrontWaves2020,dasDynamicsCrackFront2023a}.
	
	We extended the perturbation theory to the second order in $f$. The three-dimensional elastodynamic fields were analytically resolved close to the crack front and the energy release rate was computed through a self-consistent expansion~\cite{kolvinComprehensiveStudyNonlinear2024a}. Thus, $\widehat{\delta G} = \widehat{\delta G}_1[\hat{f}] + \widehat{\delta G_2}[\hat{f},\hat{f}] +\mathcal{O}(f^3)$, in which the linear part $\widehat{\delta G}_1$ is given by Eq.~(\ref{linear_dG}), and the second-order contribution is
	\begin{eqnarray}\label{deltaG2}
		&\widehat{\delta G_2}& =  2|k|P_1\lbrace \hat{f}*|k|P_1 \hat{f}\rbrace 
		-\frac{1}{2}\left\{k^2 P_1^2
		+iV\omega|k|P_2\right\}\hat{f}* \hat{f}\nonumber\\& -&\hat{f}* \left(k^2 P_1^2  - iV\omega|k| P_2\right) \hat{f} +|k|P_1\hat{f}*|k|P_1\hat{f} 
	\end{eqnarray}
	where $P_2 = P_2(\omega/|k|;V,\nu)$ is an explicit function of its arguments~(Appendix A, Eq.~(\ref{P2}),~\cite{kolvinComprehensiveStudyNonlinear2024a}), and the convolution is $(f* g)(k,\omega) = (2\pi)^{-2}\int\!\mathrm{ d}k'\,\mathrm{ d}\omega ' f(k-k',\omega-\omega ')g(k',\omega ')$. Eq.~(\ref{deltaG2}) reproduces known expressions at the limits of $k\rightarrow 0$ and $\omega\rightarrow 0$~\cite{leblondSecondorderCoplanarPerturbation2012,kolvinComprehensiveStudyNonlinear2024a,kolvinSupplementalMaterialDual}.
	
	\ik{To predict the front propagation from  Eq.~(\ref{3d_energy_balance}), the local dissipation $\Gamma$ must be specified. The fracture energy is known to depend either weakly on crack velocity, as in silica glass, or strongly as in thermoplastics and polymer gels~\cite{sharonConfirmingContinuumTheory1999,wangHowHidden3D2022,Goldman.10,hauchDynamicFractureSingle1999}.} We model the fracture energy as a product $\Gamma = \Gamma_0(V_\perp)(1+D\eta(z,x))$ of a velocity-dependent part, and a \ik{heterogeneous part whose fluctuation scales with $D$. By definition, the field $\eta$ has zero mean and unit variance. For small fluctuations,} we \ik{locally} approximate \ik{the velocity dependence} $\Gamma_0(V_\perp)\simeq \Gamma_0(V)(1+\psi (\partial_t f -V (\partial_z f)^2/2))$ where $\psi = \Gamma_0'(V)/\Gamma_0(V)$ \ik{quantifies the increase in dissipation with velocity around $V$.} We substituted the expansion $f = Df_1+D^2 f_2 + \mathcal{O}(D^3)$ in Eq.~(\ref{3d_energy_balance}) to proceed. In the zeroth order, $G_r g(V) = \Gamma_0(V)$ determines the unperturbed crack velocity~\cite{Freund.90}. The front dynamics in the first and second orders are given by the Fourier space equations
\begin{equation}\label{perturbation_hierarchy}
		\hat{f}_1 = R(k,\omega)\hat{\eta};\;\;\; \hat{f}_2 = R(k,\omega)(\widehat{\delta \Gamma}_2-\widehat{\delta G}_2[\hat{f}_1,\hat{f}_1]) \,,
	\end{equation}
	where the Green function is $R(k,\omega) = -(2|k|P_1 +i\omega \psi)^{-1}$ and 
	\begin{equation}\label{eq:delta_Gamma_2}
		\delta\Gamma_2= f_1\partial_x\eta + \psi \eta\partial_t f_1   -\psi \frac{V}{2} (\partial_z f_1)^2\,.
	\end{equation} 
    
    The formal solution given by Eqs.~(\ref{perturbation_hierarchy}) leaves the global impact of the heterogeneity implicit because of the singular behavior of $R$ at zero wavenumber and frequency. To expose the overall crack dynamics, we inspect averages of Eqs.~(\ref{perturbation_hierarchy}) over heterogeneity fields with translationally invariant correlations. Since the field $\eta$ has zero mean, the average position of the crack front remains unchanged in the first order  $\langle h(z,t)\rangle = Vt + \mathcal{O}(D^2)$ where $\langle\cdot\rangle$ denotes averaging over realizations of $\eta$. In the second order, however, crack front fluctuations contribute to the global speed of the crack. We note that $\langle \delta\Gamma_2\rangle = \gamma_2(V,\psi)$ and $\langle \delta G_2[f_1,f_1]\rangle =  g_2(V,\psi)$, where the coefficients $\gamma_2$ and $g_2$ depend on the structure of the heterogeneity field~(Appendix B). Since in general $\gamma_2-g_2\neq 0$, the front position is $\langle h\rangle = (V + v_2 D^2)t+\mathcal{O}(D^3)$ where $v_2 = (g_2-\gamma_2)/(2\pi_1+\psi)$ and $\pi_1 = \lim_{k\rightarrow 0} |k|P_1(\omega/|k|)/(i\omega)$~\cite{kolvinSupplementalMaterialDual}. Consistently with the modification of the crack speed, the nonlinear couplings contribute to the global dissipation $\langle \Gamma\rangle = \Gamma_0(V)(1 + \gamma_2 D^2 )+\mathcal{O}(D^3)$. Thus, the signs of $v_2$ and $\gamma_2$ determine whether heterogeneity assists or hinders fracture.

     \begin{figure*}[t!]
		\centering
		\includegraphics[scale = 0.9]{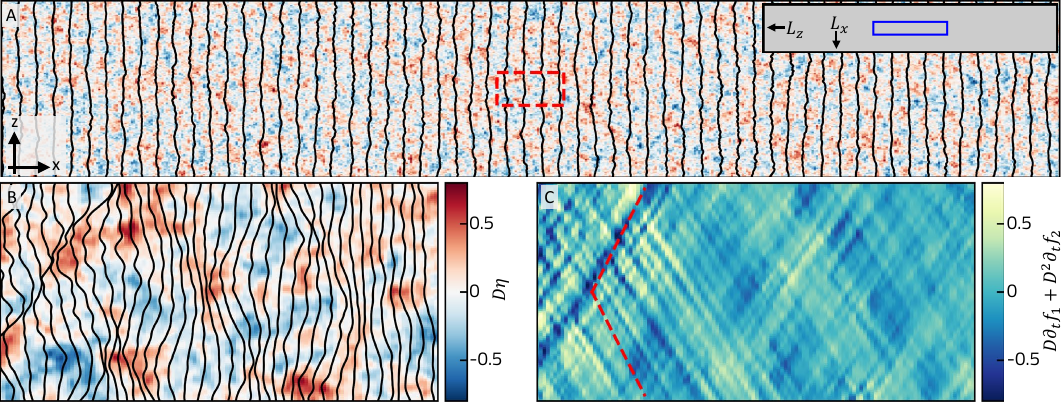}
		\caption{\textbf{Crack front dynamics across a heterogeneous toughness landscape.} (A) Fracture fronts (black lines) obtained by sampling a solution of Eqs.~(\ref{perturbation_hierarchy}) at intervals $\Delta t = 1.72$. Colors indicate the local toughness contrast $D\eta$. Inset: Geometry of the entire fracture plane including a blue rectangle corresponding to the section depicted in the main panel. $L_x = 1552, L_z = 257.8$ and $\Delta z = 0.1$.  (B) A section of the fracture plane corresponding to the dashed red rectangle in (A). Fronts are $\Delta t = 0.431$ apart. (C) The local velocity fluctuation of the fronts depicted in (B) is plotted as a function of $x=Vt$ and $z$. Red dashed lines represent the predicted front wave \ik{slopes $\pm c_f/\sqrt{c_f^2+V^2}$}~\cite{Ramanathan.97,Morrissey.00}. Solution parameters: $\nu=0.3$, $V=0.5c_R$, $D=0.2$, $\psi=0.5$.}\label{fig2}
	\end{figure*}
	
	\textit{Crack front dynamics across disordered toughness landscapes}. We evaluated Eqs.~(\ref{perturbation_hierarchy}) on a periodic rectangle $(L_x,L_z)$ \ik{comprising} an exponentially correlated Gaussian random field $\langle\eta(z,x)\eta(z',x')\rangle = \exp\left(-\sqrt{(z-z')^2+(x-x')^2}/\ell\right)$. Units were chosen such that the shear wave speed $c_s=1$ and $\ell=1$. The heterogeneity of fracture energy caused spatial and temporal crack front fluctuations (Fig.~\ref{fig2}A, Supplementary Movie 1). \ik{R}egions of increased (decreased) toughness locally slowed (accelerated) the crack front (Fig.~\ref{fig2}B). \ik{Local velocity fluctuation exhibited disturbances propagating along the crack front at the front wave speed (Fig.~\ref{fig2}C).} 
	
	\begin{figure}[b!]
		\centering
		\includegraphics[scale = 0.9]{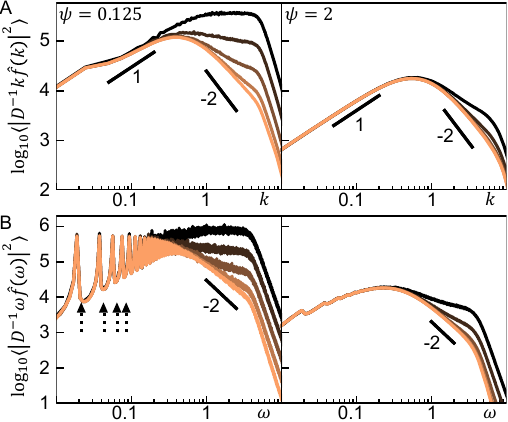}
		\caption{\textbf{(A) Static and (B) dynamic rescaled structure factors.} Colors correspond to $D$ = 0.0125 (light), 0.025, 0.05, 0.1, 0.2(dark). Curves averaged over 10 realizations of $\eta$. Black lines and adjacent numbers depict power-law scalings derived by approximating $R$ at small and large $k$~(Appendix C). Dashed arrows indicate the fundamental wave frequency $\omega_0 = 2\pi c_f/L_z$ and its harmonics. $\nu=0.3,\, V=0.5c_R$, $L_z = 257.8$, $L_x = 1552$, $\Delta z = 0.1$.}\label{fig3}
	\end{figure}
	We quantified the front geometry and dynamics via the static structure factor of the front slope  $S_k = \int \mathrm{d}\omega\; |k\hat{f}(k,\omega)|^2$ and the dynamic structure factor of the front velocity variation $S_\omega = \int \mathrm{d}k \; |\omega\hat{f}(k,\omega)|^2$. To highlight the contribution of the second-order terms, we rescaled the structure factors by the heterogeneity amplitude $D$. \ik{At small wavenumbers, $S_k$ increased monotonically as $k$~(Fig.~\ref{fig3}A). This large scale behavior is expected because of the long-range elastic interactions along the front (Appendix C).}  $S_\omega$ exhibited periodic spikes at the fundamental front wave frequency and its harmonics which were superimposed on a monotonically increasing baseline~(Fig.~\ref{fig3}B). \ik{This is an indication that elastic interactions are carried by persistent front waves at large length and time scales.} For small $D$, \ik{the structure factors peaked and then decayed approximately as $k^{-2}$ and $\omega^{-2}$. Their shapes at small $k$ and $\omega$ remained approximately independent of $D$. In parallel, stronger heterogeneity resulted in greater intermediate spectral content between the peak and the heterogeneity length and time scales $\ell$ and $\ell/V$.} The nonlinear contributions dominated the linear spectrum at $\psi\ll 1$, and produced a sub-dominant contribution for large $\psi$. Thus, nonlinear interactions are predicted to produce significant intermediate-scale roughness and temporal fluctuations compared to the linear regime~\cite{bouchaudCanCrackFront2002}.
	
	\begin{figure}[t!]
		\centering
		\includegraphics[scale = 1]{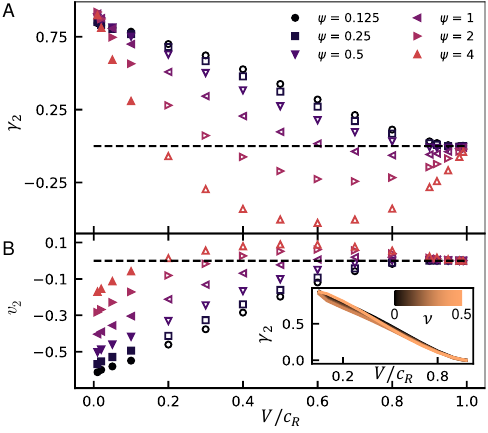}
		\caption{\textbf{Second order coefficients of (A) the fracture energy and (B) the crack velocity for disordered heterogeneity landscapes.} \ik{Eqs.~(\ref{eq:gamma_2_analytical},\ref{eq:g_2_analytical}) in Appendix C were evaluated over the Fourier domain $(k_{max} = \pi/\Delta z,\omega_{max} = \pi V/\Delta z)$ discretized with $\Delta k = 2\pi/L_z,\,\Delta \omega = 2\pi V/L_x$ for exponentially-correlated random Gaussian fields~\cite{kolvinSupplementalMaterialDual}. }  $L_z = 75.2$, $L_x = 5330.65$ (full symbols), $L_z = 1386.4$, $L_x = 288.526$ (empty symbols). Inset: Variation of $\gamma_2$ with the Poisson ratio $\nu$. $\psi = 0.125$, $L_z = 75.2$, $L_x = 5330.65$.}\label{fig4}
	\end{figure}
		
	To ascertain the impact of the heterogeneity on the global dynamics, we \ik{analytically evaluated the second-order coefficients for exponentially correlated random Gaussian heterogeneity fields. At large system sizes, the coefficients are size-independent (Appendix D, Fig.~\ref{fig:finite_size_smallpsi})~\cite{kolvinSupplementalMaterialDual,koltonUniquenessThermodynamicLimit2013}.}
	The dissipation coefficient $\gamma_2$ is positive for $\psi \ll 1$ and decreases with crack velocity (Fig.~\ref{fig4}A). Dependence on the Poisson ratio in the range $0\leq\nu\leq 0.5$ is all but completely accounted for by rescaling $V$ with $c_R$ (Fig.~\ref{fig4}B, inset). At the two limiting velocities, $V\rightarrow 0$ and $V\rightarrow c_R$, $\gamma_2$ approached $\psi$-independent limits. Upon increasing $\psi$, $\gamma_2$ decreased and attained negative values in a widening range of velocities. Thus, heterogeneity is predicted to increase dissipation in materials whose toughness is weakly velocity-dependent. In strongly velocity-dependent materials, however, dissipation is predicted to \textit{decrease} above a $\psi$-dependent velocity. Accordingly, the crack speed coefficient $v_2$ was negative for $\psi \ll 1$ (Fig.~\ref{fig4}B). For larger values of $\psi$, $v_2$ increased and attained positive values over a broad crack velocity range. At limiting velocities, $v_2$ approached a vanishing $\psi$-independent limit at $V\rightarrow c_R$ and a $\psi$-dependent value at $V\rightarrow 0$. We defined the upper bound $\tilde{D} \simeq \sqrt{\langle f_1^2\rangle/\langle (f_2-v_2 t)^2\rangle}$ to estimate the range $D\gtrsim\tilde{D}$ in which the second-order coefficients dominate. In the range of velocities $0.1c_R<V<0.9c_R$, the interval limit was $\tilde{D} \sim 20\%$ for $\psi \ll 1$, and increased to $\tilde{D}\gtrsim 100\%$ when $\psi \gg 1$ (Supplementary Fig. S1,~\cite{kolvinSupplementalMaterialDual}). These results indicate that the second-order approximation is relevant to a wide range of materials and loading conditions.	
	
	\begin{figure}[t!]
		\centering
		\includegraphics[scale = 1]{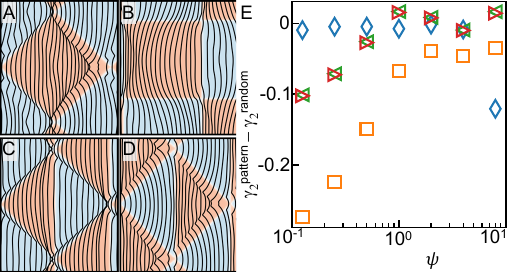}
		\caption{\textbf{Dynamic fracture of patterned planes.} Crack fronts (black lines) obtained from Eqs.~(\ref{perturbation_hierarchy}) overlaid on the toughness patterns (background colors): (A) checkered diamonds, (B) checkered squares, (C) left-pointing and (D) right-pointing triangles. Colors denote $\eta=1.2$ (light red) $\eta = -0.92$ (light blue). Side of a square in (B), 20. Time interval between crack fronts, 1.742. $\psi = 0.5$. $D=0.2$  (E) Differences between $\gamma_2$ values for the patterns (A-D) and the disordered exponentially-correlated field. System size, $L_z = 257.8$, $L_x = 1552$. All panels, $V = 0.5c_R$, $\nu=0.3$, $\Delta z = 0.1$, $\langle\eta\rangle=0$, $\langle\eta^2\rangle = 1$.}\label{fig6}
	\end{figure}
	
	\textit{Control of dynamic fracture by patterning.} Our semi-analytical framework can be used to probe the influence of toughness design on crack propagation. To shed light, we solved Eq.~(\ref{3d_energy_balance}) for four periodic patterns\ik{: reflection symmetric checkered diamonds and squares, and asymmetric triangular patterns} (Fig.~\ref{fig6}A-D, Supplementary Movies 2-5). In the diamond and triangular cases, the crack fronts exhibited large spatial and temporal gradients, whereas deformations were relatively small in the case of the squares pattern. The diamond pattern produced $\gamma_2$ that was approximately equal to the disordered case, except for large $\psi$~(Fig.~\ref{fig6}E). The checkerboard pattern exhibited much reduced $\gamma_2$. Because of the second-order terms are symmetric to the reflection $x\rightarrow -x, f\rightarrow -f$, front propagation in the asymmetric triangular patterns resulted in identical $\gamma_2$. These results indicate that dynamic fracture may be insensitive to pattern asymmetry, unlike quasi-static cracks~\cite{xiaTougheningAsymmetryPeeling2012}. \ik{Moreover, symmetric patterning may be used} to create easy crack propagation directions.
	
	\textit{Discussion.}  \ik{We developed an analytical nonlinear approximation for the dynamics of brittle crack fronts that is accurate to the second order in front perturbations. We applied this framework to predict the structure and motion of crack fronts traversing heterogeneous toughness landscapes}. We demonstrated that  nonlinear interactions \ik{between the front and the toughness landscape} result in net nonzero contributions to the global dissipation and crack speed. These effects correlated with the amplification of intermediate-scale front fluctuations in space and time. We now elucidate how nonlinearities result in toughening or detoughening and discuss open questions stemming from our work.
    
    \ik{We have seen that the second-order correction to the fracture energy, Eq.~(\ref{eq:delta_Gamma_2}), contains three contributions arising from the front translation, the velocity dependence, and the propagation along the front normal. The effects of the two first contributions stem from the short-time anti-correlation of velocity and toughness: tougher spots locally decelerate the crack and vice versa (Fig.~1). However, they have opposite impacts on \ik{dissipation}. The front translation contribution $f_1\partial_x \eta$ results from evaluating the fracture energy at the front's actual position and not at the unperturbed one.} In the limit of \ik{large systems, $\langle f_1\partial_x\eta\rangle \rightarrow - (L_x L_z)^{-1}\int \!\mathrm{d} t\mathrm{d} z\,  \eta\partial_t f_1 $, where we omitted a boundary term that vanishes as $L_x^{-1}$. The reminder is positive because front velocities and toughness variations are anti-correlated.} The second contribution, $\psi \langle  \eta \partial_t f_1\rangle $, \ik{which results from the dissipation's velocity dependence}, is negative by the same argument. The third contribution $-\frac{1}{2}\psi V\langle(\partial_z f_1)^2\rangle$,  resulting from front propagation along the local normal, is negative definite. Thus, the competition between the positive \ik{front translation term and the negative velocity-dependence terms may have toughening or weakening effects.} Moreover, the nonlinear corrections vanish as $V\rightarrow c_R$ because higher crack inertia lessens the retardance of the front by asperities. We confirmed these conclusions by an analytical calculation (Appendix B, Eq.~(\ref{eq:gamma_2_analytical})) \ik{and numerical solutions for a single circular asperity (Supplementary Fig.~S3~\cite{kolvinSupplementalMaterialDual})}. We note that though the predicted changes to the crack velocity due to heterogeneity are smaller than $V$, the broader consequences may be significant. Particularly, rapid cracks undergo microbranching~\cite{sharonConfirmingContinuumTheory1999}, macro-branching~\cite{katzavTheoryDynamicCrack2007}, and fragmentation~\cite{moulinetPoppingBalloonsCase2015} at critical subsonic crack speeds. Heterogeneity may then help to push cracks beyond the threshold.
    
How \ik{would these} predictions change for finite bodies and crack lengths~\cite{xueElastomersFailEdge2024,bazantDeterminationFractureEnergy1987}? In strip-like geometries, where $L_x,L_z\ll L_y$ the supplied energy per unit crack length is a constant~\cite{Goldman.10}. The integral dissipation $\int\!\mathrm{d} l_c \,\Gamma(z,t;V_\perp)$ over the front contour $l_c(t)$  is therefore fixed by the loading~\cite{wangHowHidden3D2022}. \ik{Disorder-generated} crack front \ik{roughness} will increase $l_c$. \ik{However, the velocity-dependence effects we described may still act to facilitate fracture. Finite-size effects may also result from loading the crack at a distance $b$ from the front. The loading} adds $\mathcal{O}(||f||/b)$ terms to $G$, which give rise to front wave dispersion in the linear order~\cite{norrisMultiplescalesApproachCrackfront2007,kolvinComprehensiveStudyNonlinear2024a}. Expanding $G$ to the second order in $f/b$ may produce terms that contribute to the effective toughness~\cite{vasoyaStudyTensileFailure2014,lebihainSizeEffectsToughening2023}. \ik{Finite length cracks may also exhibit transients that deviate from our predictions.} In heterogeneous media with $\psi=0$, front fluctuations grow linearly in time~\cite{Morrissey.00}. An inspection of $R(k,\omega)$ indicates that, for $0<\psi\ll 1$, fronts \ik{should} approach steady state at $t \sim |P_1'(c_f)|(\pi c_f \psi)^{-1} L_z$, the lifetime of the longest front wave. \ik{Time-stepping numerical solutions of Eq.~(\ref{3d_energy_balance}) may reveal the nature of such transients.}

The framework developed above may be used to explore \ik{irreversible crack dynamics. When} $V\sim\mathcal{O}(\partial_t f)$ local front arrests $V_\perp =0$ have a significant role. By transforming $\widehat{G}$ into the time domain and allowing only $V_\perp\geq 0$, fronts may be numerically evolved in time to determine the corrections to the dissipation at finite crack speed. The investigation may be extended to the depinning transition, where $G\sim\Gamma_0(V=0)$, for which tools from the theory of critical phenomena are necessary.

\phantom{The dominance of front waves when $\psi\ll 1$ gives rise to nonlinear wave-wave interactions. Since $P_2(c_f)\neq 0$~\cite{kolvinComprehensiveStudyNonlinear2024a} waves have a non-zero contribution to $G$ at the second order. When $\psi\lesssim D$, resonant wave interactions dominate the \textit{first} order terms and populate small wavelengths in $S_k$~(Fig.~\ref{fig3}A). Possibly, the hierarchy in Eqs.~(\ref{perturbation_hierarchy}) will be replaced then with a self-consistent scheme for the wave amplitudes, as in wave turbulence.}

	\begin{acknowledgments}
		The authors acknowledge enlightening discussions with the participants of the CECAM flagship workshop 3D cracks and crack stability (June 2023, Lausanne). I.K. is grateful for fruitful conversations with Sharad Ramanathan and Shmuel Rubinstein. \ik{We thank Eran Bouchbinder for a critical reading of an earlier version of the manuscript.}
	\end{acknowledgments}

	\bibliography{CFWT}

\onecolumngrid
\section*{End Matter} 
\twocolumngrid

\appendix

\counterwithin{equation}{section}
\addtocounter{equation}{-1}

\section{Appendix A: Explicit expressions for $P_1$ and $P_2$}\label{sec11}
\setcounter{equation}{0}
\renewcommand\theequation{A\arabic{equation}}

In a separate manuscript~\cite{kolvinComprehensiveStudyNonlinear2024a}, we derived the explicit expressions for the corrections to the energy release rate $\widehat{\delta G}_1(k,\omega)$ and $\widehat{\delta G}_2(k,\omega)$ that are defined in the main text.
The kernels, $P_1(\omega/|k|;V)$ and $P_2(\omega/|k|;V)$ are

\begin{align}\label{P1}
	&P_1(s;V) = -\frac{1}{2}\gamma_l\sqrt{1-\frac{\gamma_l^2 s^2}{c_l^2}}+\gamma_R\sqrt{1-\frac{\gamma_R^2 s^2}{c_R^2}}\nonumber\\
	&+\frac{1}{2}\int_{c_s}^{c_l}\frac{(s^2+V^2) (\eta^2+V^2)-2 \eta^2 V^2}{  \sqrt{\eta^2-(s^2+V^2)} \left(\eta^2-V^2\right)^2}  \Theta(\eta)\,\mathrm{d}\eta\,,
\end{align}
and
\begin{align}\label{P2}
	&P_2(s;V) = 2\frac{\gamma_R^3}{c_R^2}\sqrt{1-\frac{\gamma_R^2s^2}{c_R^2}}-\frac{\gamma_l^3}{c_l^2}\sqrt{1-\frac{\gamma_l^2s^2}{c_l^2}}+\nonumber\\
	&\int_{c_s}^{c_l}\frac{(s^2+V^2) (3\eta^2+V^2)-2 \eta^2(\eta^2+ V^2)}{ \sqrt{\eta^2-(s^2+V^2)} \left(\eta^2-V^2\right)^3}  \Theta(\eta)\,\mathrm{d}\eta\,,
\end{align}
where  $\gamma_l = 1/\sqrt{1-V^2/c_l^2}$;$\gamma_R = 1/\sqrt{1-V^2/c_R^2}$ ;  and

$$
\Theta(\eta) = \frac{2}{\pi}\arctan\left[4\sqrt{1-\frac{\eta^2}{c_l^2}}\sqrt{\frac{\eta^2}{c_s^2}-1}/(2-\frac{\eta^2}{c_s^2})^2\right]\,.
$$
$c_l$ and $c_s$ are the longitudinal and shear wave speeds respectively. The branch cuts of the square roots are defined such that $\sqrt{1-s^2}=i\,\mathrm{sign}(s)\sqrt{s^2-1}$ for $|s|>1$.

\section{Appendix B: Integral expressions for the dissipation and velocity coefficients}\label{appendix:C}
\setcounter{equation}{0}
\setcounter{figure}{0}
\renewcommand\theequation{B\arabic{equation}}
\renewcommand\thefigure{B\arabic{figure}}

In this section, we derive integral expressions for the renormalization prefactors $\gamma_2$ and $g_2$. Let us revisit the resolution of the divergence problem at $(k=0,\omega=0)$. Substituting the formal expansion $f = Df_1 + D^2f_2$ in the local energy balance Eq.~(1) we arrive at the perturbation hierarchy,
\begin{eqnarray}
	&&R(k,\omega)^{-1}\hat{f}_1 = \hat{\eta}\label{eq:first_order} \\
	&&R(k,\omega)^{-1}\hat{f}_2 = \widehat{\delta \Gamma}_2 - \widehat{\delta G}_2[\hat{f}_1,\hat{f}_1]\label{eq:second_order}\,.
\end{eqnarray}
These equations produce finite $\hat{f}_1$ and $\hat{f}_2$ everywhere except at the origin where $R(k,\omega)^{-1}\rightarrow 0$. We can gain insight into the solution at the origin by studying the averages of Eqs.~(\ref{eq:first_order},\ref{eq:second_order}). 
In the linear order, $\langle\hat{\eta}\rangle =0$ and therefore $R^{-1} \langle \hat{f}_1 \rangle= 0$ where $\langle\cdot\rangle$ denotes ensemble averaging. The family of solutions to this equation is $\hat{f}_1(k,\omega;C) = R\hat{\eta} + (2\pi)^2\delta(k)\delta(\omega) C$ where $C$ is an arbitrary displacement of the front in the crack propagation direction.  Since $G$ and the heterogeneity correlations are translationally invariant, we can set $C$ = 0 without loss of generality. Next, taking the average of Eq.~(\ref{eq:second_order}) we have 
$$
R^{-1} \langle \hat{f}_2 \rangle= \langle\widehat{\delta \Gamma}_2\rangle - \langle\widehat{\delta G}_2[\hat{f}_1,\hat{f}_1]\rangle\,.
$$
Let us focus on the first term in $\langle\widehat{\delta \Gamma}_2\rangle$ which is $\langle\hat{f}_1*(V^{-1}i\omega\hat{\eta})\rangle$.
Writing explicitly,
\begin{align}
	&\langle\hat{f}_1*(V^{-1}i\omega\hat{\eta})\rangle = \frac{1}{(2\pi)^2V}\int\!\mathrm{d}k'\mathrm{d}\omega'\, \nonumber \\
	&R(k-k',\omega-\omega') i\omega' \langle \hat{\eta}(k-k',\omega-\omega')\hat{\eta(k',\omega')}\rangle\,,\nonumber
\end{align}
and applying the Weiner-Khinchin theorem to compute $\langle\hat{\eta}(k-k',\omega-\omega')\hat{\eta(k',\omega')}\rangle = (2\pi)^2\delta(k)\delta(\omega)S(p')/V$, we have $\langle\hat{f}_1*(V^{-1}i\omega\hat{\eta})\rangle = \delta(k)\delta(\omega) V^{-2}\int\!\mathrm{d}k'\mathrm{d}\omega'\, R(-k',-\omega') i\omega'S(p')$. Generalizing this procedure to the rest of the terms, we see that $\langle\widehat{\delta \Gamma}_2\rangle = (2\pi)^2\delta(k)\delta(\omega) \gamma_2$ where 
\begin{equation}\label{eq:gamma_2_analytical}
	\gamma_2 = \int_\mathcal{Q} \frac{S(p)}{\pi^2}\left[\frac{1-\psi V}{V^2}\,\omega \,\mathrm{Im} \{R\} - \frac{\psi}{2}k^2 |R|^2\right]\,.
\end{equation}
and $\mathcal{Q} = \{(k,\omega)\in \mathbb{R}^2: k>0  \;\mathrm{and}\; \omega>0\}$.
Similarly, $\langle\widehat{\delta G}_2[\hat{f}_1,\hat{f}_1]\rangle = (2\pi)^2\delta(k)\delta(\omega) g_2$ where
\begin{equation}\label{eq:g_2_analytical}
	g_2 =\int_\mathcal{Q}\frac{S(p)}{\pi^2 V}\big[k^2 (|P_1(s)|^2 - P_1(s)^2 )+iV\omega k P_2(s)\big]|R|^2\,.
\end{equation}
and $s = \omega/k$. Due to their distinct origins, $g_2$ and $\gamma_2$ need not be equal in general, and do not equal zero individually as can be assessed from the diverse terms that comprise them. We therefore rewrite the average of Eq.~(\ref{eq:second_order}) as
\begin{equation}
	R(k,\omega)^{-1}\langle\hat{f}_2\rangle = (2\pi)^2\delta(k)\delta(\omega) (\gamma_2 - g_2)\,,
\end{equation}
with the solution $\langle\hat{f}_2\rangle = (2\pi)^2\delta(k)(i\omega)^{-1}\delta(\omega) (\gamma_2 - g_2)/(\psi+2\pi_1)$. We can now inverse Fourier transform to obtain $\langle f_2\rangle = v_2 t$ where
\begin{equation}\label{v_2_analytical}
	v_2 =\left( g_2 - \gamma_2\right)/\left(\psi +2\pi_1\right)\,.
\end{equation}

To evaluate the integrals in Eqs. (\ref{eq:gamma_2_analytical},\ref{eq:g_2_analytical}), the $(k,\omega)$ space was discretized at steps of size $\Delta k = 2\pi/L_z$ and $\Delta \omega = 2\pi V /L_x$, The maximum wavelength was $k_{max} = \pi/\Delta z$ and the maximum frequency was $\omega_{max} = \pi V /\Delta z$.
The integrals were then computed in NumPy using a trapezoidal approximation.
Note that $\mathrm{Im} \{R\}$ is non-negative, which means that $\gamma_2$ takes positive values for $\psi\ll 1$, and may become negative for large enough $\psi$.

\section{Appendix C: Scaling of the static structure factor}
\setcounter{equation}{0}
\setcounter{figure}{0}
\renewcommand\theequation{C\arabic{equation}}
\renewcommand\thefigure{C\arabic{figure}}

To elucidate the factors contributing to the shape of the static structure factor in the first-order, we analyze the Green function $R(k,\omega) = -(2|k|P_1 +i\omega \psi)^{-1}$ of the linear problem. The expression can be simplified in two limits. For $\psi \ll 1$, the Green function is sharply peaked around the front wave dispersion relation $\omega = c_f k$. We can then linearize the kernel $P_1(\omega/|k|)\simeq \mp P_1'(c_f)(\omega / |k| \mp c_f)$ since $P_1(c_f)=0$ and obtain
\begin{equation}\label{eq:R_smallpsi}
	R(k,\omega) \simeq [\mp 2 |k| P_1'(c_f)(\omega / |k| \mp c_f) - i\psi \omega)]^{-1}
\end{equation}
where the signs are $-$ or $+$  for $\omega >0$ and $\omega<0$ respectively.
Conversely, for $\psi \gg 1$, the front wave peak is obliterated and the response is centered around $(k=0,\omega=0)$. We can then approximate
\begin{equation}\label{eq:R_largepsi}
	R(k,\omega)\simeq -(2|k|P_1(0) + i\psi\omega)^{-1}\,.
\end{equation}

Using these approximations, we evaluate the linear front fluctuation term $\hat{f}_1(k,\omega)$ for an exponentially correlated random Gaussian heterogeneity field and derive expressions for the static structure factor of the front slope  $S^1_k = \int \mathrm{d}\omega\; |k\hat{f_1}(k,\omega)|^2$. Since the linear term is given by $\hat{f}_1 = R\hat{\eta}$, we find that 
$$
S^1_k\simeq (\pi V)^{-1} \int_0^\infty \mathrm{d}\omega |kR|^2 S(p)
$$
where $S(p)$ is the structure factor of the heterogeneity field and $p = \sqrt{k^2+(\omega/V)^2}$.
For $\psi\ll 1$, $S(p)$ is slowly varying at the front wave dispersion ridge, while $R$ is sharply peaked. Then, $S(p)\simeq S\left(k\sqrt{1+(c_f/V)^2}\right)$ and using Eq.~(\ref{eq:R_smallpsi}) we obtain
$$
S^1_k\simeq\frac{1}{V \psi P_1'(c_f)c_f} |k| S\left(k\sqrt{1+(c_f/V)^2}\right)\,.
$$
This spectrum scales as $S^1_k\sim k$ at small wavenumbers and as $S^1_k\sim k^{-3}$ at large scales since $S(p)\sim \mathrm{const.}$ for $p\ll \ell^{-1}$ and $S(p)\sim p^{-3}$ for $p\gg \ell^{-1}$.
Similarly, for $\psi\gg 1$, the Green function is concentrated around $\omega =0$ and it is possible to approximate $S(p)\simeq S\left(k\right)$ in that region. The application of Eq.~(\ref{eq:R_largepsi}) then results in the expression
$$
S^1_k\simeq\frac{1}{4 V \psi P_1(0)} |k|S(k)\,.
$$
In this limit, $S^1_k\sim k$ for $k\ll \ell^{-1}$ and $S^1_k\sim k^{-3}$ for $k\gg \ell^{-1}$.

\section{Appendix D: Size dependence of the coefficients}
\setcounter{equation}{0}
\setcounter{figure}{0}
\renewcommand\theequation{D\arabic{equation}}
\renewcommand\thefigure{D\arabic{figure}}

\begin{figure}[b]
	\centering
	\includegraphics[scale = 1]{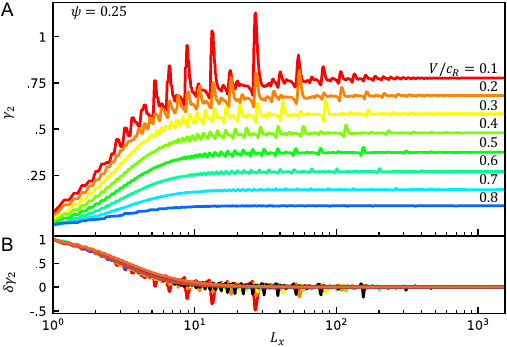}
	\caption{\textbf{The effect of finite system size on the renormalization prefactor $\mathbf{\gamma_2}$.} (A) The fracture energy prefactor as a function of fracture plane length for a range of crack velocities  (B) Collapse of  curves in (A) and Fig. S2~\cite{kolvinSupplementalMaterialDual} by normalizing $\delta\gamma_2 = (\gamma_2(L_x) - \gamma_2(1551))/ (\gamma_2(1) - \gamma_2(1551))$.  Solution parameters, $\nu=0.3$, $L_z = 257.8$, $\Delta z = 0.1$.}\label{fig:finite_size_smallpsi}
\end{figure}

The renormalized dissipation exhibited finite-size dependence~\cite{koltonUniquenessThermodynamicLimit2013,lebihainSizeEffectsToughening2023}. To investigate, we numerically computed $\gamma_2(L_x;V,\psi)$ for a fixed $L_z$. $\gamma_2(L_x)$ increased monotonically with $L_x<10\ell$ where $\ell$ is the heterogeneity correlation length (Fig.~\ref{fig:finite_size_smallpsi}A,B). At $L_x\sim 10\ell$, $\gamma_2$ approached an asymptotic size-independent value. Through a linear transformation, all the functions $\gamma_2(L_x;V,\psi)$ approximately collapsed on a single master curve, showing that the asymptotic length scale $L_x\sim 10\ell$ was independent of $V$ and $\psi$ (Fig.~\ref{fig:finite_size_smallpsi}B). The manner of the asymptotic approach, however, changed with the parameters. For $\psi \geq 1$, $\gamma_2$ approached the asymptote smoothly, without appreciable deviations from it for $L_x>20\ell$. For $\psi\ll 1$, $\gamma_2$ approached the asymptote while exhibiting spike-like fluctuations that persisted until $L_x>10^3$. The amplitudes of the isolated spikes increased inversely with the crack velocity. Since  $\gamma_2$ was evaluated through Eq.~(\ref{eq:gamma_2_analytical}), where ensemble averaging is analytically performed, there is no stochastic component that could give rise to fluctuations. Instead, the finite-size spikes are expected to arise from front wave propagation in a finite geometry.

\end{document}